\documentclass[twocolumn,preprintnumbers,aps,amsmath,amssymb]{revtex4}
\usepackage{graphicx}% Include figure files
\usepackage{dcolumn}% Align table columns on decimal point
\usepackage{bm}% bold math
\usepackage{epsfig}
\begin{document}

%\preprint{APS/123-QED}

\title{Theory of the highly viscous flow}% Force line breaks with \\

\author{U. Buchenau}
 \email{buchenau-juelich@t-online.de}
\affiliation{%
Forschungszentrum J\"ulich GmbH, J\"ulich Centre for Neutron Science (JCNS-1) and Institute for Complex Systems (ICS-1),  52425 J\"ulich, Germany}%

\date{December 28, 2020}% It is always \today, today,

\begin{abstract}
The recent theoretical treatment of irreversible jumps between inherent states with a constant density in shear space is extended to a full theory, attributing the shear relaxation to structural Eshelby rearrangements involving the creation and annihilation of soft modes. The scheme explains the Kohlrausch exponent close to 1/2 and the connection to the low temperature glass anomalies. A continuity relation between the irreversible and the reversible Kohlrausch relaxation time distribution is derived. The full spectrum can be used in many ways, not only to describe shear relaxation data, but also to relate shear relaxation data to dielectric and bulk relaxation spectra, and to predict aging from shear relaxation data, as demonstrated for a very recent aging experiment. 
\end{abstract}

\maketitle

\section{Introduction}

An increasing part of our knowledge of glasses and the glass transition is due to numerical results \cite{olig,kob,corei,cavagna,le1,manning,le2,mizuno,iso,wang,le4,corein,le6,edan}. The pioneering work of Schober, Oligschleger and Laird \cite{olig} revealed the string character of the soft vibrations and low barrier relaxations in a soft sphere glass. Seven years later, the same kind of motion was found in the viscous flow of a binary glass former \cite{kob}. Though the authors did not realize it, the result shows that the flow motion is identical with the one of the soft modes responsible for the low temperature glass anomalies, the main point of the present paper.

The low temperature glass anomalies are well described by the soft potential model \cite{ramos,herbert,abs}, an extension of the tunneling model \cite{phillips} to include soft vibrations and low barrier relaxations. New numerical results confirm three soft potential model predictions, namely the existence of a strong fourth order term in the mode potential \cite{le1}, a density of soft vibrational modes increasing with the fourth power of their frequency \cite{le1,manning,le2,mizuno,wang}, and a density of low barrier relaxations increasing proportional to the power 1/4 of the barrier height \cite{le4}. A strong decrease of the number density of the soft modes with decreasing glass temperature was established \cite{wang} using the new swap mechanism \cite{swap} to cool numerical glass formers to very low freezing temperatures. 

Another crucial numerical finding \cite{corei,corein,edan} is the unstable core of the soft modes ("soft spot"), stabilized by the surroundings. The mode displacement is accompanied by a long range elastic distortion describable in terms of an elastic dipole tensor in the mode center which is essentially a planar shear dipole \cite{le6,edan}. But the eigenvector outside the unstable core reflects also a strong interaction with other soft spots in the immediate neighborhood \cite{edan}.

The present paper devises a theory of the highly viscous flow combining the new numerical insight with long-known experimental facts. The most important of these is the extrapolated divergence of the viscosity at the Kauzmann temperature, where the excess entropy of the undercooled liquid over the crystal extrapolates to zero \cite{angell}. The excess mean square displacement of the liquid over the crystal also disappears at the Kauzmann temperature \cite{bzr,hansen1,hansen2}, indicating that the soft spots disappear there, in accordance with one of the new numerical results \cite{wang}.

The treatment starts with a recent exact result \cite{bu2018a,bu2018b} on the flow process, describing the relaxation spectrum of irreversible Eshelby transitions \cite{eshelby} between inherent states with a constant density in the five-dimensional shear space around the Maxwell time $\eta/G$ ($\eta$ viscosity, $G$ shear modulus).

The present paper adds the reversible part, describing both reversible and irreversible Eshelby transitions in terms of the creation and annihilation of the soft modes responsible for the low temperature glass anomalies, and explaining the Kohlrausch exponent in terms of the increasing number $N_s$ of participating soft modes with increasing barrier height. A continuity relation connects the reversible Kohlrausch part to the irreversible spectrum around the Maxwell time.

The next Section II summarizes the previous work \cite{bu2018a,bu2018b} on the irreversible shear relaxation, analyzes the free energy of an unstable core with two stable structural minima in the undercooled liquid, attributes the Eshelby barrier of both irreversible and reversible transitions to the sum of the crossing energies of unstable cores, and derives the full spectrum. Section III compares the theory to experimental data, first to oscillations in the frozen Kohlrausch tail in the glass phase, then to glass transition data, not only to shear relaxation, but to compressibility, dielectric and aging data as well. Section IV discusses and concludes the paper.

\section{Full shear relaxation spectrum}

\subsection{Summary of previous work}

Consider a more or less spherical region of several hundred atoms containing $N_s$ soft spots. It is reasonable to define a dimensionless shear misfit $e$ of the region in such a way that $e^2$ is the elastic misfit energy in units of $k_BT$. In thermal equilibrium, the states $e$ in the five-dimensional shear misfit space have an average energy of 5/2 $k_BT$ in the normalized distribution
\begin{equation}\label{pe}
	p(e)=\frac{1}{\pi^{5/2}}e^4\exp(-e^2).
\end{equation}
The prefactor corrects the one of eq. (3) in the previous paper \cite{bu2018a}, a mistake which does not invalidate the results of the paper.

As shown there, the jump rate for the Eshelby \cite{eshelby} transition from $e_0$ to $e$ gets a factor $\exp((e_0^2-e^2)/2)$ from the difference in the shear misfits. With this, the state $e_0$ has the escape rate
\begin{align}\label{e0r}
	r_0=N_er_V\frac{8\pi^2}{3}\int_0^\infty\exp((e_0^2-e^2)/2)e^4de \nonumber\\
	=4\sqrt{2}\pi^{5/2}N_er_V\exp(e_0^2/2),
\end{align}
where $N_e$ is the density of stable structural alternatives to the initial state in the five-dimensional $e$-space \cite{bu2009} and $r_V$ is the jump rate for the barrier height $V$ between two states with the same elastic misfit energy. $N_e$ increases with the size of the region, because an additional particle adds structural excess entropy.

Averaging the escape rate of eq. (\ref{e0r}) over all occupied states of eq. (\ref{pe}), one finds the average escape rate
\begin{equation}
	\overline{r}=32\pi^{5/2}r_VN_e.
\end{equation}

This average escape rate equals the single jump rate $r_V$ already at the relatively low density 
\begin{equation}\label{ns}
	N_e=\frac{1}{32\pi^{5/2}},
\end{equation}
where the jumps become irreversible.

The previous paper \cite{bu2018a} shows that for the critical region size, the average escape rate is
\begin{equation}
r_c=\frac{1}{\tau_c}=\frac{G}{8\eta},	
\end{equation}
where $G$ is the high frequency shear modulus and $\eta$ is the viscosity.

The normalized decay time distribution in the barrier variable $v=\ln(\tau/\tau_c)$ is \cite{bu2018b}
\begin{equation}\label{pt}
	l_{irrev}(v)=\frac{1}{3\sqrt{2\pi}}\exp(v^2)\left(\ln(4\sqrt{2})-v\right)^{3/2}.
\end{equation}

In this picture, the last regions to relax are the unstrained ones with the relaxation time $4\sqrt{2}\tau_c$, a factor of forty five longer than the Maxwell time $\tau_M=\eta/G$. 

Eq. (\ref{pt}) for the relaxation time distribution of the irreversible processes turned out to be able to describe dynamic heat capacity data with the Maxwell time from shear data for four different glass formers, among them two hydrogen bonded substances \cite{bu2018a,bu2018b}.

Another previous result, which will be needed for the calculation of the Kohlrausch tail in Section II. D, concerns the effectivity difference between irreversible and reversible Eshelby transitions for the shear relaxation. A reversible relaxation is a factor 0.4409/2=0.22045 less effective than an irreversible one \cite{bu2018b}.

Note that the strong decrease of the structural excess entropy with decreasing temperature \cite{angell} implies a decrease of the density of inherent states in the five-dimensional shear space, therefore an increase of the size of the critical region, with a divergence at the Kauzmann temperature.

\subsection{Soft modes at the glass transition}

%%%%%%%%%%%%%%%%%%%%% begin figure %%%%%%%%%%%%%%%%%%%%%%%%%%%%%%%%%%%%%
\begin{figure}[b]
\hspace{-0cm} \vspace{0cm} \epsfig{file=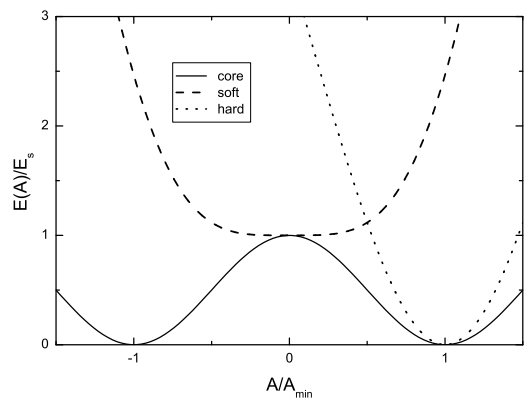,width=7 cm,angle=0} \vspace{0cm} \caption{Mode potentials in units of the creation energy $E_s$ of the soft mode as a function of the normal coordinate $A$: continuous line core potential, dashed line full soft mode potential including the harmonic restoring forces from the surroundings, dotted line potential for the mode if the minimum of the outside restoring forces coincides with the minimum of the core potential at $A_{min}$ (see text).}
\end{figure}
%%%%%%%%%%%%%%%%%%%%% end figure %%%%%%%%%%%%%%%%%%%%%%%%%%%%%%%%%%%%%%%

Let us first visualize a soft mode in flowing surroundings, where the zero point $a_0$ of the outer stabilizing forces for the unstable core can flow away from the soft mode value $a_0=0$.

The unstable soft core of the mode \cite{corein,corei,le6,edan} is a local saddle point configuration between two minima, lower than the saddle point in energy by the creation energy $E_s$ of the soft mode and held in place by the surroundings. The continuous line in Fig. 1 shows the core potential $E_c(A)$ alone as a function of the normal coordinate $A$, with the neighboring minima at $\pm A_{min}$, described by a cosine function. The dashed line is the sum of the core potential and the harmonic restoring forces from the surroundings, together bringing the restoring force of the soft mode down to zero and leaving only the stabilizing quartic potential term.

If the surroundings are able to flow, the zero point $a_0$ of the harmonic outside force can shift. The dotted curve in Fig. 1 shows the mode potential for the case that the zero point coincides with the core potential minimum at $A_{min}$. 

The effective frequency $\omega_s$ of the soft mode at the temperature $T$ in terms of the soft potential parameter $W$ is given by the mean square displacement in the purely quartic potential
\begin{equation}\label{uq}
	<x^2>=\frac{\Gamma(3/4)^2}{\sqrt{2}\pi}\left(\frac{k_BT}{W}\right)^{1/2}
\end{equation}
leading to
\begin{equation}\label{oms}
	\hbar\omega_s=\frac{2^{3/4}\pi^{1/2}}{\Gamma(3/4)}(k_BT)^{1/4}W^{3/4}\approx2.43(k_BT)^{1/4}W^{3/4}.
\end{equation}

From this equation, the soft mode frequency for ten glasses \cite{ramos} at their glass temperature turns out to lie close to their boson peak, so even at the glass temperature the soft mode gains the vibrational entropy $S_s$.

For the creation energy $E_s$, values between $k_BT_g$ and 6 $k_BT_g$, with an average value of 2.5 $k_BT_g$, were found numerically \cite{edan}. Together with the negative contribution $-TS_s$ to the free energy, this explains the strong increase of their number density \cite{bzr,hansen1,hansen2} above $T_g$.

For a given $E_s$, it is easy to calculate the free energy of a mode for any value of $a_0$ between zero and $A_{min}$ numerically. This allows to determine the average mean square displacement and the distribution of the modes over the $a_0$-values. For a creation energy of 2.5 $k_BT_g$, one finds the average excess energy 0.56 $k_BT_g$ per mode over the one at $a_0=A_{min}$ and a vibrational excess entropy $T_gS_s$ of 0.13 $k_BT_g$. Their ratio is 0.23, not too far from the ratio 0.28 between excess vibrational and excess structural entropy found experimentally \cite{se} in selenium.

\subsection{Eshelby barriers}

Let us now return to the Eshelby transition of a large region of several hundred atoms discussed in Section II. A, from one stable structure to another sheared one. According to the central postulate of the present paper, the motion from one minimum to the other is a displacement of a number $N_s$ of soft modes. Let us restrict this number to those which cross their saddle point $a_0=0$ in the course of the motion. 

%%%%%%%%%%%%%%%%%%%%% begin figure %%%%%%%%%%%%%%%%%%%%%%%%%%%%%%%%%%%%%
\begin{figure}[t]
\hspace{-0cm} \vspace{0cm} \epsfig{file=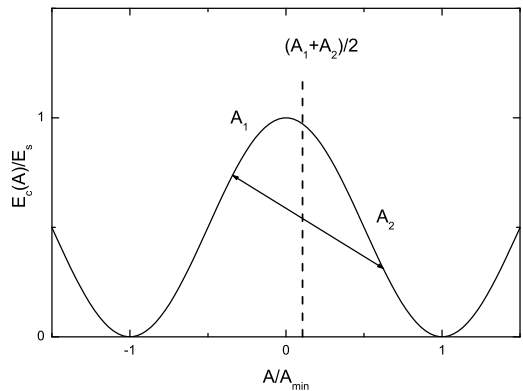,width=7 cm,angle=0} \vspace{0cm} \caption{A soft core mode participating in an Eshelby jump changes the normal coordinate saddle point distance from the negative value $A_1$ to the positive value $A_2$ and contributes the energy $E_c((A_1+A_2)/2)-E_c(A_1)$ to the Eshelby barrier. For a jump in the opposite direction, the contribution is $E_c((A_1+A_2)/2)-E_c(A_2)$.}
\end{figure}
%%%%%%%%%%%%%%%%%%%%% end figure %%%%%%%%%%%%%%%%%%%%%%%%%%%%%%%%%%%%%%%

In order to determine the height $E_b$ of the Eshelby barrier, one has to sum up the appropriate average over the $N_s$ contributions. For example, the mode in Fig. 2 starts in one structural minimum with its $a_0$ at the negative value $A_1$ with the core potential $E_c(A_1)$ and crosses over to the positive value $A_2$, so its unstable core potential has the value $E_c((A_1+A_2)/2)$ at the top of the barrier. In that case, the difference of these two values is the contribution of the mode to the barrier height. Note that if $A_1$ is close to the saddle point and $A_2$ is close to $A_{min}$, the contribution is negative.

If starting position and end position are exchanged, then the difference of $E_c((A_1+A_2)/2)$ and $E_c(A_2)$ counts.

To determine the full barrier, one has to sum over all $N_s$ contributing soft modes, with the weight of $A_1$ determined by its free energy at the given temperature, derived in the preceding subsection II. B. Taking the average creation energy 2.5 $k_BT_g$ found in a numerical glass \cite{edan}, the numerical integration over all possible combinations of $A_1$ and $A_2$ yields an average barrier contribution of $\Delta E_b=1.46k_BT_g$ for a crossing soft mode.

With this result, by adding one crossing mode the Eshelby barrier increases by $\Delta E_b$ and one adds two structural possibilities, increasing the barrier density by a factor of two. The corresponding increase of the barrier density with increasing height is described by the Kohlrausch exponent $\beta$
\begin{equation}
	\beta=\frac{k_BT_g\ln{2}}{\Delta E_b}.
\end{equation}

For $\Delta E_b=1.46k_BT_g$, this gives $\beta=0.48$, close to the average value found experimentally in many metallic glasses \cite{met}.

The finding of a Kohlrausch exponent close to 1/2 for all glass formers \cite{bohmer,albena} at $T_g$ shows that the creation energy of the soft modes is always about 2.5 $k_BT_g$. It determines the number density of tunneling states freezing in at the glass temperature, a number density which has been found to be essentially universal in all glasses \cite{phillips,hunklinger}.

\subsection{Continuity relation}  

At a fixed temperature in the undercooled liquid, the number $N_s$ of crossing soft modes on the critical region is fixed, fixing the energy barrier for its percolating (and therefore irreversible) Eshelby transitions and its size.

The Eshelby transitions of smaller regions, responsible for the Kohlrausch tail, are no longer percolating and are therefore reversible back-and-forth jumps, though with a finite lifetime. Their decay rate distribution is given by eq. (\ref{pt}) for the critical region.    

The cutoff of the reversible Kohlrausch relaxations at $\tau_c$ can be calculated by integrating this decay function of the irreversible states, from the smallest relaxation times up to the relaxation time $\tau$ of a given relaxation in the Kohlrausch tail. One finds that the cutoff is well approximated by the Fermi function $F(v)=1/(1+\exp(1.19v))$ in the barrier variable $v=\ln(\tau/\tau_c)$.

The Kohlrausch $\beta$ implies that one has a barrier density $l(v)\propto\exp(\beta v)$ with increasing barrier height. As one approaches $\tau_c$, a fraction $1-F(v)$ decays irreversibly, leaving a fraction $F(v)$ to contribute to the reversible Eshelby relaxations.  

The sample has the viscous shear compliance $8/G$ within the terminal relaxation time $\tau_c=8\eta/G$ \cite{bu2018a,bu2018b}. One must have enough irreversible decay states below the longest decay time $4\sqrt{2}\tau_c$ to achieve this, a condition which allows to fix the normalization factor of the Kohlrausch barrier density numerically.

It follows that the complex shear compliance $J(\omega)$ is
\begin{equation}\label{jom}
	GJ(\omega)=1+\int_{-\infty}^\infty \frac{l_K(v)dv}{1+i\omega\tau_c\exp(v)}-\frac{i}{\omega\tau_M},
\end{equation}
with the barrier density $l_K(v)$ of the reversible Kohlrausch processes
\begin{equation}\label{lv}
	l_K(v)=(1+0.115\beta-1.18\beta^2)F(v)\exp(\beta v).
\end{equation}

The factor in the first parenthesis on the right side of eq. (\ref{lv}) results from the normalization condition, taking into account that the average irreversible decay of a state yields a factor 2/0.4409 more compliance response than the average reversible back-and-forth jump \cite{bu2018b}.

The normalized barrier density of the irreversible processes is given in eq. (\ref{pt}). The total barrier density of reversible and irreversible Eshelby processes is
\begin{equation}\label{ltot}
	l_{tot}(v)=l_0(8l_{irrev}(v)+l_K(v)),
\end{equation}
where $l_K(v)$ is the Kohlrausch tail of eq. (\ref{lv}) and the normalization factor $l_0\approx 0.085$ depends on $\beta$, though not much.

\section{Comparison to experiment}

To begin with, the validity of the concept is supported by undeniable periodic wiggles in the long time shear relaxation of a metallic glass at room temperature \cite{atzmon}. The oscillations have a period corresponding to a barrier difference $\Delta V=0.064$ eV. In the present interpretation, this barrier difference is due to the addition of another crossing soft string mode. With the glass temperature of 529 K \cite{alcu}, it corresponds to 1.4 $k_BT_g$, close to the value derived here.

Very recently, similar oscillations have been discovered \cite{atti} in old sound attenuation measurements \cite{pino} in selenium, with a period corresponding to 0.036 eV, in terms of the glass temperature $T_g=307$ K again 1.4 $k_BT_g$.

The good fit of the scheme to shear relaxation data has already been shown earlier \cite{bu2018a,bu2018b}, though with the reversible fraction $f_r$ of the shear relaxation as free parameter. Here, it is determined by eq. (\ref{lv}), leaving only $G$, $\eta$ and $\beta$ as free parameters. Fig. 3 (a) shows the excellent fit of the very accurate $G(\omega)$-data \cite{tina} for the vacuum pump oil DC704, a molecular glass former, in terms of eq. (\ref{jom}).

The successful quantitative connection between reversible and irreversible shear processes in eq. (\ref{ltot}) can be exploited to understand what one sees at the $\alpha$-relaxation peak of other techniques. In the dynamic specific heat, one sees only the irreversible processes. This was already demonstrated for four glass formers in the two previous papers \cite{bu2018a,bu2018b} and is understandable, because a back-and-forth jump can only store a small amount of heat.

The situation is different for density and dielectric relaxation. If irreversible and reversible processes are both Eshelby processes of the same kind, the total barrier density in dielectric, density and shear relaxation should be the same.

This is indeed the case. It is demonstrated in Fig. 3 (b), which displays dielectric and bulk relaxation data taken on the same sample in the same cryostat \cite{tina}, described with the same parameters as the shear data in Fig. 3 (a) plus one additional relaxation time.

%%%%%%%%%%%%%%%%%%%%% begin figure %%%%%%%%%%%%%%%%%%%%%%%%%%%%%%%%%%%%%
\begin{figure}[t]
\hspace{-0cm} \vspace{0cm} \epsfig{file=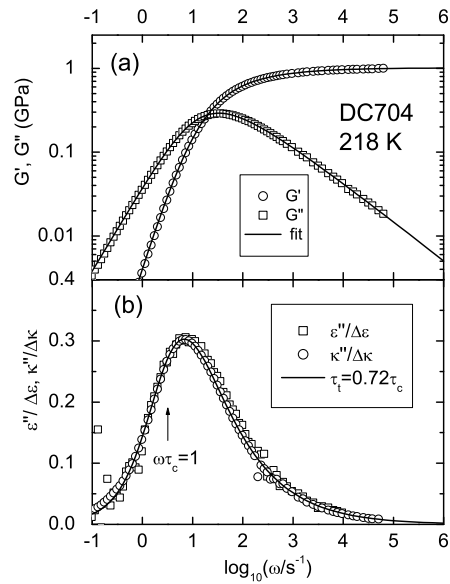,width=6 cm,angle=0} \vspace{0cm} \caption{(a) Measurement \cite{tina} of $G(\omega)$ in the vacuum pump oil DC704; theory with $G$=1.03 GPa, $\eta$=0.038 GPas, $\beta=0.47$ (b) Fit of dielectric and adiabatic compressibility data for the same sample in the same cryostat with $\tau_t=0.72\tau_c$.}
\end{figure}
%%%%%%%%%%%%%%%%%%%%% end figure %%%%%%%%%%%%%%%%%%%%%%%%%%%%%%%%%%%%%%%

But both the molecular reorientation and the adiabatic density relaxation terminate before $4\sqrt{2}\tau_c$, both at approximately the same terminal relaxation time $\tau_t=0.72\tau_c$. The total barrier density of eq. (\ref{ltot}) has to be multiplied with $\exp(-\tau(v)/\tau_t)$ and the normalization factor $l_0$ has to be recalculated.

The two peaks in Fig. 3 (b) happen to lie close together, but this is mere coincidence. In the next example, another vacuum pump oil, PPE, a chain of five connected phenyl rings, the dielectric peak lies lower and the adiabatic compressibility peak lies higher, independent of the temperature \cite{jakobsen}, so they no longer coincide.

One concludes that both the density and the dielectric polarization equilibrate before the last memory of the shear disappears. If the process stops before the shear fluctuations end, it must be because the faster part of the shear fluctuations have already managed to equilibrate the density and rotate every molecule.

%%%%%%%%%%%%%%%%%%%%% begin figure %%%%%%%%%%%%%%%%%%%%%%%%%%%%%%%%%%%%%
\begin{figure}[t]
\hspace{-0cm} \vspace{0cm} \epsfig{file=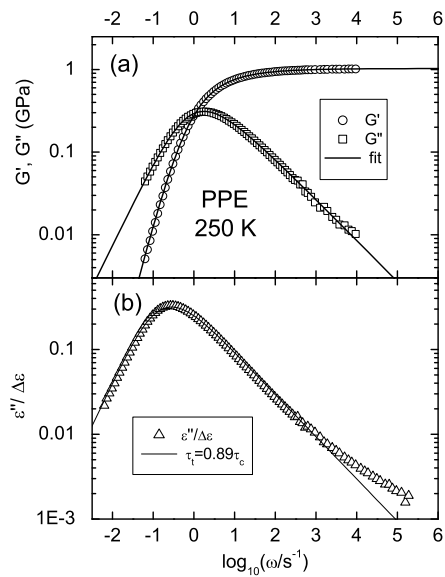,width=6 cm,angle=0} \vspace{0cm} \caption{(a) Measurement \cite{tina} of $G(\omega)$ in another vacuum pump oil, PPE; theory with $G$=1.03 GPa, $\eta$=0.725 GPas, $\beta=0.48$ (b) Fit of dielectric data for the same sample in the same cryostat with $\tau_t=0.89\tau_c$.}
\end{figure}
%%%%%%%%%%%%%%%%%%%%% end figure %%%%%%%%%%%%%%%%%%%%%%%%%%%%%%%%%%%%%%%

Fig. 4 (a) shows the theory for the $G(\omega)$-data \cite{tina} of PPE. Again, the agreement is excellent. Fig. 4 (b) displays a fit of dielectric data from the same sample in the same cryostat with the same parameters and a cutoff at $\tau_t=0.89\tau_c$. The dielectric data show a clear excess wing over the Kohlrausch tail, which is not contained in the theory.

If one forces the excess wing of the dielectric data on a fit of the shear data, the quality of the description deteriorates. In fact, the logarithmic curvature of the Kohlrausch tail of the shear data averages to zero within an error bar which is a factor two smaller than the logarithmic curvature of the dielectric data. The result indicates that the excess wing in PPE is due to processes which, unlike the Kohlrausch processes, do not couple strongly to the shear.

In another example, propylene carbonate \cite{pc}, the Kohlrausch tails of both the shear and the dielectric data have a visible logarithmic curvature. But the logarithmic curvature of the shear data is again more than a factor of two smaller than the one of the dielectric data.

%%%%%%%%%%%%%%%%%%%%% begin figure %%%%%%%%%%%%%%%%%%%%%%%%%%%%%%%%%%%%%
\begin{figure}[t]
\hspace{-0cm} \vspace{0cm} \epsfig{file=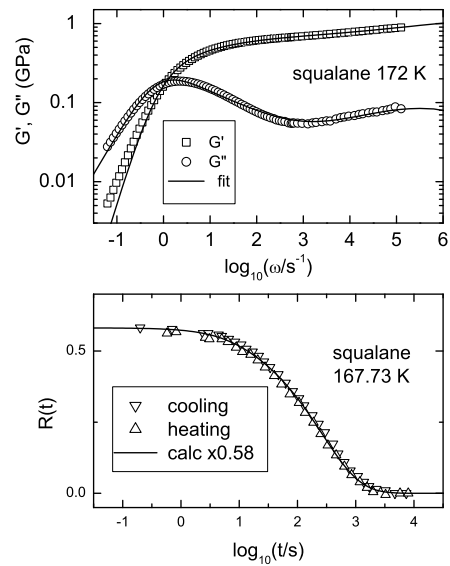,width=6 cm,angle=0} \vspace{0cm} \caption{(a) Measurement \cite{nibo} of $G(\omega)$ in squalane; theory with $G$=1.167 GPa, $\eta$=0.396 GPas, $\beta=0.363$ and a secondary relaxation gaussian at 0.273 eV with a full width at half maximum of 0.42 eV and the amplitude 0.07  (b) Measured \cite{ag2020} and calculated aging curves in squalane at 167.73 K.}
\end{figure}
%%%%%%%%%%%%%%%%%%%%% end figure %%%%%%%%%%%%%%%%%%%%%%%%%%%%%%%%%%%%%%%

Propylene carbonate and glycerol \cite{bu2018b} are examples where the theoretical shear description of the present paper is not perfect: One has to shift the cutoff for the reversible states to $3\tau_c$, to considerably longer times, thus introducing an additional parameter. Then one retrieves a good fit of the shear data.

This indicates that propylene carbonate and glycerol belong to those glass formers where some memory remains after the thermal equilibration time $\tau_c$, though on a much smaller scale than in polymers \cite{adachi,alexei} and mono-alcohols \cite{mono,catalin}, where one finds marked peaks at times much longer than $\tau_c$.

On the other hand, one finds a multitude of examples where the present theory of the shear relaxation works very well. Among them is the ionic glass former CKN \cite{bier}, vitreous silica \cite{mills}, a window glass \cite{win}, many metallic \cite{schro,met} and many molecular \cite{nibo,mckenna,maggi} glass formers. 

One can also include a secondary relaxation peak, adding three free parameters for its description. This is shown in the last example, squalane, in Fig. 5 (a), for the $G(\omega)$-data \cite{nibo} at 172 K. Since there are also data at 174, 176 and 178 K, one can make a Vogel-Fulcher extrapolation to $\tau_c=188.7$ s at 167.73 K, the temperature of a very recent aging experiment \cite{ag2020}. Then one can compare the measured equilibration with the prediction derived from eq. (\ref{ltot}). This is shown in Fig. 5 (b). Apart from the starting value 0.58, there is no adapted parameter in the calculated curve, which lies nicely between the cooling and the heating data and corroborates the conclusion \cite{ag2020} that there is a terminal relaxation time (in the present description at forty five times the Maxwell time), even though the relaxation time distribution goes over many decades. 

\section{Discussion and conclusions}

\subsection{Eshelby barriers}

The central assumption of the present treatment is that the Eshelby barrier between two stable structures of a region is the sum of the crossing energies of the local shear modes which change their minimum in the transition. The questions raised by this assumption need to be discussed.

The first question is whether the creation energy (the enthalpy) or the free energy of the local mode determines its contribution to the barrier height.

This question is answered by the finding of the undiminished frozen Kohlrausch tail in vitreous silica in the glass phase \cite{philmag2002} at half the glass temperature, demonstrating that one deals with well-defined enthalpy barriers. Evidence for an undiminished Kohlrausch tail has also been found in the glass phase of selenium \cite{atti}, another glass former with no secondary relaxation peak. Together with the experimental evidence for the weak oscillations in the frozen Kohlrausch tail in the metallic glass \cite{atzmon} and in selenium \cite{atti}, the findings exclude a strong entropy contribution to the Eshelby barriers.

One concludes that the vibrational entropy of the soft modes is important for their concentration at the glass temperature, where they freeze in, but does not enter their contribution to the Eshelby barriers. 

A second issue is the elastic interaction between different local soft shear modes with large displacements, for which there is numerical evidence \cite{edan}. These strong interactions need to be taken into account.

But these interactions are crucial to enable the formation of the second independent stable structure of the region. A given crossing local shear mode in an Eshelby transition finds a new minimum because the other crossing local modes have changed the shear distortion of its surroundings. The consideration shows that the mode is exposed to strong forces when it passes its saddle point, explaining why the vibrational entropy is not relevant for the Eshelby barrier.

The present treatment explains the Kohlrausch stretching of the viscous flow in terms of the reversible Eshelby barriers. If their barrier height increases proportional to the number of crossing soft modes, one can explain the universal finding \cite{met,bohmer,albena} of a Kohlrausch $\beta$ close to 1/2 in terms of a universal creation energy of the unstable core of the soft modes close to 2.5 $k_BT_g$, the value recently found in a numerical glass \cite{edan}. This is compatible with the finding of a universal density of tunneling modes in glasses at low temperatures \cite{phillips,hunklinger}. 

\subsection{Mean square displacements}

According to eq. (\ref{uq}), the mean square displacement of a soft mode, though being twenty times larger than the one of a transverse zone boundary mode, increases only with $T^{1/2}$ at the glass temperature.

An amusing consequence is that the apparently perfectly harmonic boson peak in liquid germania \cite{uqge} is not harmonic at all, but appears harmonic because the number of soft modes increases proportional to $T^{1/2}$ in the undercooled liquid. This interpretation of the data \cite{uqge} is supported by a slight decrease with increasing temperature of the neutron scattering intensity below the boson peak and a slight increase above, in accordance with the frequency proportional to $T^{1/4}$ of eq. (\ref{oms}).

In selenium, there is a measurement \cite{zorn} of the excess mean square displacement of the undercooled liquid over the crystal from the glass temperature up to the melting point. With equs. (\ref{uq}) and (\ref{oms}), one can calculate the number of soft modes per atom, which extrapolates to zero at the Kauzmann temperature. One finds one soft mode per twenty five atoms at the glass temperature and one soft mode per six atoms at the melting temperature, an increase of the density of soft modes by about a factor of four.

As pointed out in the previous publication \cite{zorn}, the variation of the viscosity over fifteen orders of magnitude between 290 and 600 K is well described by a proportionality of $1/\log{\eta}$ to the excess mean square displacement, suggesting the same flow mechanism from the aging glass up to the hot liquid.

But the present paper limits itself to a quantitative description of the Kohlrausch stretching. A quantitative description of the fragility is beyond its scope, though the experimental findings \cite{bzr,hansen1,hansen2} suggest it might be a worthwhile task. 

\subsection{Conclusions}

To conclude, the previous quantitative analysis of the irreversible shear relaxation processes is extended to a full theoretical description of reversible and irreversible shear relaxation in undercooled liquids. The Kohlrausch behavior with a Kohlrausch exponent close to 1/2 is explained in terms of the soft modes responsible for the low temperature glass anomalies. The energy barrier against the shear transformation of an Eshelby region is attributed to the sum of crossing energies of those local shear modes which have to displace and find a new minimum in order to reach the new stable structure. 

The theory works in network, ionic, metallic and molecular glass formers, but seems to require some as yet unknown modification to be able to describe hydrogen bonding substances. The description gives approximately the right Kohlrausch $\beta$ in all glass formers. It provides the long-sought connection between glass transition and low temperature glass anomalies. It has a well-defined terminal relaxation time in good agreement with a recent aging experiment. It is consistent with the equality of Vogel-Fulcher and Kauzmann temperatures in the extrapolation of the experimental data of many glass formers.

The data that supports the findings of this study are available within the article.

\end{document}